# THE X-RAY CLUSTER NORMALIZATION OF THE MATTER POWER SPECTRUM

J. PATRICK HENRY,[1] AUGUST E. EVRARD,[2] HENK HOEKSTRA,[3] ARIF BABUL,[3] AND ANDISHIEH MAHDAVI[3]

## ABSTRACT

The number density of galaxy clusters provides tight statistical constraints on the matter fluctuation power spectrum normalization, traditionally phrased in terms of $\sigma_8$, the root mean square mass fluctuation in spheres with radius $8h^{-1}$ Mpc. We present constraints on $\sigma_8$ and the total matter density $\Omega_{m0}$ from local cluster counts as a function of X-ray temperature, taking care to incorporate and minimize systematic errors that plagued previous work with this method. In particular, we present new determinations of the cluster luminosity – temperature and mass – temperature relations, including their intrinsic scatter, and a determination of the Jenkins mass function parameters for the same mass definition as the mass – temperature calibration. Marginalizing over the 12 uninteresting parameters associated with this method, we find that the local cluster temperature function implies $\sigma_8 \, (\Omega_{m0}/0.32)^{\alpha} = 0.86 \pm 0.04$ with $\alpha = 0.30 \, (0.41)$ for $\Omega_{m0} \leq 0.32$ ($\Omega_{m0} \geq 0.32$) (68% confidence for two parameters). This result agrees with a wide range of recent independent determinations, and we find no evidence of any additional sources of systematic error for the X-ray cluster temperature function determination of the matter power spectrum normalization. The joint WMAP5 + cluster constraints are: $\Omega_{m0} = 0.30^{+0.03}_{-0.02}$ and $\sigma_8 = 0.85^{+0.04}_{-0.02}$ (68% confidence for two parameters).

*Subject headings:* cosmological parameters – cosmology: observations – galaxies: clusters: general – Large-scale structure of universe – X-rays: galaxies

## 1. INTRODUCTION

Present data concerning the structure in and evolution of the universe are well described by a theory using just six principal parameters (Tegmark et al., 2006; Komatsu et al., 2008 among many others). These parameters are: $\Omega_{m0}$ and $\Omega_{b0}$, the present total matter and baryon densities with respect to critical density respectively; h, the present value of the Hubble parameter in units of 100 km s$^{-1}$ Mpc$^{-1}$; $\sigma_8$, the present rms total matter fluctuations in spheres of $8 \, h^{-1}$ Mpc radius; $n_s$, the primordial power spectrum spectral index; and $\tau$, the optical depth to the last scattering. This minimal model, which we assume in this paper, sets several other parameters to specific values, most notably the spatial curvature = 0 (flat) and the equation of state parameter of the dark energy = -1 (cosmological constant).

The normalization of the matter fluctuation power spectrum P(k) comes from $\sigma_8$ via

[1] Institute for Astronomy, 2680 Woodlawn Drive, Honolulu, HI 96822; henry@ifa.hawaii.edu
[2] Departments of Physics and Astronomy and Michigan Center for Theoretical Physics, University of Michigan, Ann Arbor, MI 48109-1040
[3] Department of Physics and Astronomy, University of Victoria, Victoria, BC V8W 3P6, Canada

$$\sigma_8 = \left[ \int_0^\infty dk \, \frac{k^2}{2\pi^2} P(k) \left[ \frac{3j_1(8k)}{8k} \right]^2 \right]^{1/2} \quad (1)$$

or $\sigma_8 \approx [P(0.172 \, h \, Mpc^{-1})/3879 \, h^{-3} \, Mpc^3]^{1/2}$. (Peacock, 1999, eqs. [16.13] and [16.132]) where $j_1$ is the Spherical Bessel function of the first kind, order one and k is the spatial wavenumber. The shape of P(k) is determined by $\Omega_{m0}$, $\Omega_{b0}$, h, and $n_s$ (Eisenstein & Hu, 1998).

The number density of clusters of galaxies at a given epoch is sensitive to some of these cosmological parameters while the evolution of the number density is sensitive to others. The sensitivity to h, $\Omega_{b0}$, $n_s$ and τ is weak or nonexistent; is moderate to $\Omega_{m0}$; and is strong to $\sigma_8$. There has thus been much effort expended trying to exploit this sensitivity to measure $\sigma_8$. Generally the values from the cluster method have been lower than those coming from other techniques. For example, Hetterscheidt et al. (2007) compile cluster and weak lens cosmic shear determinations since 2001 and find the 2002 – 2006 averages of $\sigma_8$ assuming $\Omega_{m0}$ = 0.3 are 0.728 ± 0.035 and 0.847± 0.029, respectively. Spergel et al. (2003) find 0.92 ± 0.10 from the WMAP first year data. This discrepancy, along with the perceived complicated cluster X-ray gas physics, has led to a slow acceptance of the cluster-based values. This situation was summarized in the 2006 Final Report of the Dark Energy Task Force as: "…the prediction of [cluster] counts is subject to substantial uncertainties in the baryonic physics…This method is the one for which our forecasts are least reliable, due to this large astrophysical systematic effect." (Albrecht et al., 2006)

It has not always been emphasized that the strong dependence of cluster number density on $\sigma_8$ comes with an equally strong dependence on systematic effects. Perhaps the largest systematic uncertainty is the relation between the cluster mass and a more easily observable proxy, called the mass – observable relation. Henry (2004; H04 hereafter, Figure 9 and Table 4, which gives earlier cluster determinations of $\sigma_8$) shows that much of the scatter among reported cluster $\sigma_8$ measurements is simply due to the assumed mass-temperature relation normalization.

Two recent developments motivated this paper. First, the three and five year WMAP results for $\sigma_8$ are 0.76 ± 0.05 (Spergel et al., 2007) and 0.796 ± 0.036 (Dunkley et al., 2008). Although the changes with respect to WMAP1 are of marginal statistical significance, the new values of $\sigma_8$ do agree with the historical average from X-ray clusters quoted above. As summarized in Tegmark et al. (2006) Section IV.B.1, this reduction comes from a reduction of the best fit values of τ, $\Omega_{m0}h^2$, and $n_s$, resulting from improved modeling of noise and foregrounds, better statistics, and an improved analysis procedure. Second, as we explain in Section 3, the three *a priori* best methods of calibrating the cluster mass – temperature relation, masses from weak gravitational lensing, masses from the equation of hydrostatic equilibrium corrected for non-thermal pressure support applied to the X-ray gas of likely virialized objects and numerical hydrodynamic simulations, are, for the most part, consistent. The agreement indicates the calibration of this crucial relation is on a firmer basis than what was previously possible.

We here present an updated measurement of $\sigma_8$ derived from a local sample of cluster temperatures chosen to minimize systematic errors but including their effects in the analysis. This update uses a new mass – temperature (M – T) calibration with temperatures from the same source as the sample (or put on that scale) and a new determination of the mass function with the same definition of cluster mass as the M − T calibration. We measure ourselves, or use the WMAP5 measurements of, all twelve uninteresting parameters needed for the analysis, including their estimated errors. We marginalize over all these uninteresting parameters when reporting our constraints in the $\sigma_8$ - $\Omega_{m0}$ plane. Readers not interested in these technical details may skip to Section 5 where we give the resulting cosmological constraints and Section 6 for a comparison of them to those provided by other methods.

## 2. CLUSTER SAMPLE

Present cluster determinations of $\sigma_8$ are already systematics limited. So there is no need to decrease the statistical errors on $\sigma_8$ by increasing the sample size or redshift range beyond that of previous samples if the goal is to measure it. Larger samples of calibrators external to the statistical sample used to do cosmology can be useful when measuring various relations needed to derive the cosmological constraints. Larger statistical samples may be useful when trying to identify its residual systematic error by breaking it into subsamples with various properties. Our goal here is to minimize systematic errors on the determination of $\sigma_8$. To that end we use a local sample that minimizes evolutionary effects. We also want only massive objects since groups may not be scaled down versions of clusters. For example Sanderson et al. 2003 find that the luminosity – temperature relation steepens considerably for objects with temperatures less than 2 keV. Sun et al. (2008) suggest that the difference between groups and hotter clusters is mainly due to differences between their cores. The sample we chose is HIFLUGCS (Reiprich & Böhringer, 2002). This sample is X-ray selected and X-ray flux limited from the ROSAT All-Sky Survey, but with fluxes redetermined from ROSAT Position Sensitive Proportional Counter (PSPC) pointed observations for 75% of the sample. It covers 8.14 steradians with fluxes in the 0.1 to 2.4 keV band $f_{200}(0.1,2.4) > 2.0 \times 10^{-11}$ erg cm$^{-2}$ s$^{-1}$ ($f_{200}$ is defined below). We made two additional cuts: redshift ≤ 0.2 for a local sample and temperature ≥ 3 keV for massive objects.

There are 48 objects meeting all these criteria. Their average z is 0.0551. Forty-five clusters have single temperature MEKAL model fits of ASCA data to the entire cluster (i.e. without excluding any cool cores) from Horner (2001). The temperatures of the remaining 3 objects (A1656, ZwCl1215 and A1644) are derived from Ikebe et al. (2002) after regressing Horner's temperatures against theirs for the 45 objects. This procedure yields temperatures 3% lower than the original Ikebe et al. values. The average fractional statistical 68% confidence temperature error of all 48 objects is 2.2%.

The HIFLUGCS catalog presents, among other things, the PSPC count rate in channels 52 – 201 (approximately 0.5 – 2.0 keV) within a specified outer radius that is different for each object, z, Hydrogen column density towards the source, and the beta index and core radius of a beta model spatial profile. We used all these parameters plus the temperature to derive the absorption-free flux within $r_{200}$, where $r_{200}$ is the radius within which the average density of the cluster is 200 times the critical density at the redshift of the cluster. Using the definitions of $M_{200}$, $r_{200}$, and $M_{500}$, $r_{500}$, and the relation between $M_{500}$ and kT determined below, $r_{200}$ as a function of temperature is $[15/8\pi \, M_{200}/M_{500} \, A_{MT} \, (kT)^{\alpha_{MT}}/500/\rho_c(z)]^{1/3}$. Using the Navarro, Frenk, and White (1995) mass profile with c = 5 to find $M_{200}/M_{500}$ = 1.479 and specifying $\alpha_{MT}$ = 3/2 with the corresponding $A_{MT}$, yields:

$$r_{200} = 2.77 \pm 0.02 \, h_{70}^{-1} \, \text{Mpc} \, [kT/10 \, \text{keV}]^{1/2} /E(z). \qquad (2)$$

with $h_{70}$ the present value of the Hubble parameter in units of 70 km s$^{-1}$ Mpc$^{-1}$, $\Omega_{m0} = 0.3$ and $E(z) = [\Omega_{m0} (1 + z)^3 + 1 - \Omega_{m0}]^{1/2}$. We interpolate or extrapolate the beta model from the specified outer radius to $r_{200}$. On average for the HIFLUGCS objects in our sample the outer radius extends to 0.887 of $r_{200}$. The flux within $r_{200}$, $f_{200}$, is on average 4.5% higher than the HIFLUGCS flux and this factor varies from 5.1% to 4.4% as $\Omega_{m0}$ varies from 0.05 to 0.5. That is, this correction is small, independent of h and very nearly independent of $\Omega_{m0}$. The average fractional statistical 68% confidence flux error of all 48 objects is 2.0%.

The virial theorem implies that temperature is a useful mass proxy. In massive simulated halos, the dark matter velocity dispersion scales with mass in a manner that is independent of cosmology and is well approximated as a power-law relation with fixed log-normal scatter (Evrard et al., 2008). This general form is also seen in the matter temperature of simulations that include gas, irrespective of the detailed baryonic physics treatment employed (Borgani et al., 2004; Balogh et al. 2006; Kravtsov et al., 2006). This behavior suggests that the gas physics that determines the temperature may be simple and that the parameterization of the M – T relation may not be sensitive to small changes in cosmology.

Balanced against these desirable properties is the added uncertainty of the selection function. As outlined above, the HIFLUGCS clusters are flux – selected, not temperature – selected. Equations (4) and (5) of H04 show how to determine the temperature selection function from the flux selection function. The procedure is as follows.

The luminosity and redshift may be converted to flux in the usual way yielding the solid angle surveyed in which a cluster with these properties could have been detected.

$$\Omega(f_{200}(E_1, E_2)) = \Omega(L_{200}(\text{bol}), z) = \Omega\left(\frac{L_{200}(\text{bol}) \, BF(E_1, E_2, kT(L))}{4\pi \, D_L^2(\Omega_{m0}, z) \, k(E_1, E_2, z, kT(L))}\right) \quad (3)$$

Here $L_{200}$(bol) is the bolometric luminosity within $r_{200}$, BF is the band fraction that gives the fraction of the bolometric luminosity that is in the energy band $E_1$ to $E_2$, $D_L$ is the luminosity distance with h = 0.7 and $\Omega_{m0} = 0.3$ for specificity, the same as we used for $r_{200}$, and k is the k correction. All quantities on the right hand side of (3) are known, once a cluster luminosity – temperature relation is specified, which we do below. While the flux selection function is cosmology independent, the selection function of luminosity and redshift is not.

In order to obtain the temperature selection function, we must convert the luminosity to a temperature. Key to this conversion is of course the luminosity - temperature relation, which we determine from our sample. We derive the bolometric luminosities from the fluxes as described above. Dimensional analysis suggest power-law behaviors for the mean luminosity and temperature as a function of mass and redshift, and hence a power-law luminosity – temperature relation (Nord et al., 2008; Kaiser, 1986). Power-law behavior seems justified empirically. If the relation is in fact not a power law, then the intrinsic scatter we derive would be over estimated producing a more ramp-like jump in the temperature selection function coming from over smoothing the step function flux selection function. There is no effect outside the jump region thus a small effect overall. So we fit a power law of the form $L_{44}$(bol) = $A_{LT}$ $(kT)^{\alpha LT}$ where the bolometric luminosity is in units of $10^{44}$ erg s$^{-1}$ and the temperature is in units of keV. We use the BCES bisector for the fit, which incorporates errors in both variables plus intrinsic scatter (Akritas & Bershady, 1996), see the Appendix A for more details. Readers who prefer a different temperature pivot (i. e. the temperature divided by a number other than unity, 5 is sometimes quoted in the literature) should consult Appendix B for a simple procedure to transform our results to their preferred pivot. Table 1 gives the best fit values and information on their covariance matrix. Figure 1 shows the error ellipse for the two fitting parameters. Figure 2 compares the best fit with the data.

This sample exhibits the usual good L – T relation but with a large intrinsic scatter. The intrinsic scatter is log normal (e.g. Novicki, Sornig & Henry, 2002), so we report in Table 1 the scatter of log(L) at constant kT, $\sigma_{\log(L)T}$. If there were no scatter, then we could simply replace $L_{200}$(bol) in equation (3) with a unique temperature, thereby obtaining a selection function of temperature and redshift. Because the band fraction and k correction are weak functions of temperature for the small range of temperatures considered here, the scatter in the L - T relation produces almost no scatter for them. Thus to obtain $\Omega(kT,z)$ we must average over the possible $\Omega(L_{200}(bol),z)$s at each temperature, weighted by the probability of obtaining them.

$$\Omega(kT,z) = \int d(\log L)\, \Omega(L,z) \frac{\exp\left\{-\left[\log(A_{LT}(kT)^{\alpha_{LT}} D_L^2(h,\Omega_{m0},z)/D_L^2(0.7,0.3,z)) - \log L\right]^2 / 2\sigma^2_{\log(L)T}\right\}}{\sqrt{2\pi\sigma^2_{\log(L)T}}} \quad (4)$$

where we have generalized the L – T relation to $L_{44}$(bol) = $A_{LT}(kT)^{\alpha_{LT}}$ $[D_L(h,\Omega_{m0},z)/D_L(0.7,0.3,z)]^2$. The luminosity distance dependence converts from the h = 0.7, $\Omega_{m0}$ =0.3 cosmology in which the $A_{LT}$ and $\alpha_{LT}$ are determined to any other cosmology.

Stanek et al. (2006) and Nord et al. (2008) point out that both the shape and scatter of the observed L – T relation from a flux-limited sample may be different from a mass-limited sample. The degree and direction of the difference depends on the covariance of the L and T variations at fixed mass. A non-zero covariance can enhance or suppress the probability that an object of a given mass survives the flux cut. Although we derive our L – T relation from a flux-limited sample, we expect these effects to be small for it. First, clusters in the simulations summarized in Stanek et al. (2006) exhibit weak covariance, only 2%. Second, the flux cut that defines our HIFLUGCS sample (2 x $10^{-11}$ erg cm$^{-2}$ s$^{-1}$) is a factor of ~7 higher than the flux limit of the underlying ROSAT All-Sky Survey from which it was drawn (3 x $10^{-12}$ erg cm$^{-2}$ s$^{-1}$) so the log of their ratio is 3.3 $\sigma_{\log(L)T}$. The last two terms of equation A11 from Vikhlinin et al. (2008) show that this effect is then 0.1% of the purely statistical likelihood given by the first term of A11, which is what we used to determine our L – T relation. Thus this effect is completely negligible for our application.

## 3. MASS – TEMPERATURE RELATION

We need an L – T relation because clusters are selected by flux, not temperature. Analogously, we need an M – T relation because the theory is expressed in terms of mass, not temperature. To calibrate this relation we need a sample of clusters with known masses and temperatures. The classic technique to estimate cluster masses is from observations of the radial velocities of its member galaxies. The mass may be estimated in several ways but there are the oft - discussed inherent unknowns associated with each. Girardi et al (1998) and Rines & Diaferio (2006) give short summaries of these techniques. The Jeans method requires the unknown galaxy orbital distribution (since the velocity dispersion across the line of sight is not known) or the unknown mass distribution, often assumed to be the same as the galaxies (mass follows light). The virial theorem method integrates the Jeans equation but again requires the unknown form of the mass distribution. Both of the preceding methods require that the cluster is in dynamic equilibrium. Finally the caustic method (Diaferio 1999), which is related to and not independent of the Jeans method, does not require dynamic equilibrium but still needs the galaxy orbital and total mass distributions. Information on the galaxy orbital distribution can come from measuring the galaxy velocity kurtosis, but the mass estimate still requires an assumption on the mass distribution (see Łokas et al, 2007). We attempted to calibrate the M – T relation from the thirteen $M_{500}$ caustic masses of Rines & Diaferio (2006) with Horner (2001) temperatures using the procedure described below. The measured dispersion is $0.80 \pm 0.22$, a factor of 6 larger than what we find below. We attribute this large dispersion to the ambiguities inherent with galaxy velocities used to measure masses and do not pursue this method further.

There are three other ways to obtain cluster masses: 1. Weak lensing observations, 2. X-ray observations of relaxed clusters assuming virial equilibrium with corrections for non-thermal pressure support and 3. Numerical simulations that calculate both the mass and temperature. Although historically the three methods have not always agreed, we show here consistency among them, suggesting convergence to the true solution. This result ameliorates the largest systematic uncertainty in using cluster number density measurements to constrain cosmological parameters. We fit a power law to characterize the M – T relation: $E(z)^h M_{500,15} = A_{MT} (kT)^{\alpha_{MT}}$, where $M_{500,15}$ is the mass inside a spherical overdensity of 500 with respect to the critical density at the redshift of the cluster in units of $10^{15}$ M⊙ and kT is in keV. Here we use $\Omega_{m0} = 0.3$ to be specific. In Appendix A we describe how we performed the fits.

### 3.1. *X-ray Temperatures, Weak Lensing Masses*

In principle this method should be the most reliable as the masses and temperatures are determined completely independently. We used the masses of the seventeen clusters in Table 3 of Hoekstra (2007), as revised by Mahdavi et al. (2008) that have X-ray temperatures from Horner (2001). This is the same temperature source as was used for the L – T relation. Three Hoekstra clusters are not included (MS 1231+15, A209 and A383) because they have no Horner temperatures. This sample has an average redshift of 0.289. Figure 3 shows the error ellipse for the two fitting parameters.

### 3.2. *X-ray Temperatures, X-ray Masses*

Care must be exercised with this method because to use it the objects must be in hydrostatic equilibrium and both the temperature and surface brightness must be spatially resolved. One or more of these requirements were not met in many previous analyses. An additional complication for our particular application is the mass interior to a radius is proportional to the temperature at that radius, which in turn is nearly a constant fraction of the average temperature when that radius encloses a density that is a constant fraction of the critical density (e.g. Pratt et al., 2007). Thus correlating mass measured this way against average temperature is correlating something strongly dependent on average temperature against average temperature and this dependence may modify the M – T relation that is finally derived.

Table 4 of Vikhlinin et al. (2006) contains thirteen clusters, all with X-ray temperatures and ten with measured $M_{500}$ masses. These objects have a very regular X-ray morphology, indicating that they are likely in virial equilibrium. The sample of ten has an average redshift of 0.109. The temperatures reported by Vikhlinin et al. were measured with Chandra data and excluded the inner 70 $h_{72}^{-1}$ kpc. Only seven of the ten clusters with mass measurements have Horner (2001) temperatures. We calculated a weighted average ratio of the Vikhlinin et al. $T_{spec}$ to the Horner T for the eight of the thirteen clusters that have Horner temperatures, finding 1.083 ± 0.008 (68% confidence). This slight positive bias is expected, since Horner includes any cool core gas in his measurement while Vikhlinin et al. (2006) do not. We divided the Vikhlinin et al. temperatures of the sample of ten by this factor to place them on the Horner scale.

Two further corrections are needed, both derived from the simulations described in Section 3.3. The first arises because relaxed clusters likely have different masses than the same temperature cluster that has experienced a recent merger. The statistical correction factor for including merging clusters in the sample can be derived by comparing the masses of merging clusters to the same temperature relaxed clusters. We find this a factor of 1.122 ± 0.055 (68% confidence) by comparing $A_{MT}$s for All z, all clusters to All z, relaxed clusters in Table 2 of Kravtsov, Vikhlinin & Nagai (2006). The second correction is for the effects of non-thermal pressure support even in relaxed looking clusters. From Table 2 of Nagai, Vikhlinin & Kravtsov (2007), for $M_{tot}(<r_{est})$, z = 0, $r_{500c}$, Relaxed clusters, this correction is a factor of 1.242 ± 0.042 (68% confidence error on the mean for a sample of 21 objects). That is relaxed looking clusters are 24% more massive than deduced from assuming hydrostatic equilibrium. There is observational support for the latter correction. Mahdavi et al. (2008) find it to be 1.28 ± 0.15 comparing weak lensing masses with X-ray hydrostatic masses. The product of these two corrections is 1.394 ± 0.083. Note that Table 2 of Nagai, Vikhlinin & Kravtsov (2007) gives this total correction for all clusters as 1.339 ± 0.075, which agrees with what we calculated from the more circuitous route that begins with relaxed looking clusters because the data start there. This agreement may reflect the results from simulations that most massive halos are close to hydrostatic equilibrium, at least within $r_{500}$, whether they look relaxed or not (Evrard, Metzler & Navarro, 1996; Rasia et al. 2006). Figure 3 shows the error ellipse for the two fitting parameters after making the corrections described in the previous two paragraphs.

### 3.3. *Numerical Simulations of Temperatures and Masses*

The physics of most of the cluster, that outside the inner ~100 kpc, may be very simple. Collisionless dark matter dominates the total mass, the hot gas is nearly completely ionized by collisions, and the radiation comes from optically thin thermal bremsstrahlung. Consequently there has long been the hope that numerical hydrodynamic simulations could accurately calculate observational properties of clusters. Kravtsov et al. (2006) report one of the most recent such calculations. It includes dissipationless dark matter dynamics, gas dynamics, star formation, metal enrichment due to Type Ia and II supernovae, self-consistent advection of metals, metallicity-dependent radiative cooling, thermal feedback from supernovae, stellar winds and stellar mass loss, and UV heating due to the ionizing background. It does not include AGN feedback, cosmic rays or magnetic fields. Further, the reported temperatures of the simulated clusters are what an actual instrument (the ACIS on Chandra) would have observed, not the emission measure weighted temperature often employed previously.

We used the sixteen clusters with simulated masses and temperatures from Table 1 of Kravtsov et al. (2006). The sample was constructed to have a redshift of 0.0. Kravtsov et al. (2006) measure their temperatures excluding a central region that is a fixed fraction of $r_{500}$ (0.15), rather than a fixed metric radius (70 $h_{72}^{-1}$ kpc) as do Vikhlinin et al. (2006). Kravtsov et al. determine that their temperatures are 0.97 of the latter's or 1.051 ± 0.008 (68% confidence) times Horner's. No corrections are needed to the simulation masses since they are known from summing all simulation particles within the appropriate radius, which is applicable to all clusters relaxed or unrelaxed. Figure 3 shows the error ellipse for the two fitting parameters.

### 3.4. *Joint Fit*

The individual fits agree within their 68% confidence limits, as is shown in Figure 3. The $A_{MTS}$ agree with each other to ±3%; the $\alpha_{MTS}$ agree with each other and with the self-similar slope also to ±3%. The three fits are not quite independent. The weak lensing and X-ray mass samples have one object in common, A2390. The X-ray masses are increased by two factors coming from the simulations. Nevertheless, the samples are nearly independent, so we performed a joint fit. Table 1 gives the best fit values and information on their covariance matrix. Figure 4 show the error ellipse. We compare this best fit to the data in Figure 5. Readers who prefer a different temperature pivot should consult Appendix B for a simple procedure to transform our results to their preferred pivot.

We used a different M – T relation in our previous work (H04 and references therein), one relating the virial mass and temperature via h $M_{v,15}$ ~ $(\beta_{TM} kT)^{3/2}$. For comparison, the relation between the two definitions is

$$\beta_{TM} = 8.0 \left[ A_{MT} \frac{M_v}{M_\Delta} \right]^{2/3} \left[ \frac{\Delta_{vc}}{18\pi^2} \right]^{1/3} \quad (5)$$

where $M_v$ is the virial mass, $\Delta$ = 500, $\Delta_{vc}$ = $\Omega_m \Delta_v$ is the overdensity within the virial radius with respect to the critical density and $\Delta_v$ is with respect to the background density (see Henry, 2000 for the latter). For $\Omega_{m0}$ = 0.3 and z = 0.0551, $\Delta_{vc}$ = 104.81. A Navarro, Frenk, & White (1995, NFW) mass profile with c = 5 and the same cosmology and redshift yields $M_v/M_{500}$ = 1.832. So the $A_{TM}$ we find here implies $\beta_{TM}$ = 1.07 ± 0.04.

Again we will vary both h and $\Omega_{m0}$ when deriving our cosmology constraints, so need to generalize the M – T relation from the cosmology used to obtain it to an arbitrary cosmology

$$E(z) h M_{500,15} = A_{MT} (kT)^{\alpha MT} \frac{D_L(1, \Omega_{m0}, z)}{D_L(1, 0.3, z)}. \quad (6)$$

The dependence on the luminosity distance is exact for the X-ray masses, but is actually $D_l D_s/D_{ls}$ for weak lensing, where the Ds are angular diameter distances to the lens (the cluster), source and from lens to source respectively. We can use luminosity distances here since the distances in equation (6) enter only as ratios, and the angular diameter and luminosity distances differ only by $(1 + z)^2$. $D_s/D_{ls}$ only varies by ±5% for the average source and lens redshifts of the calibrators as $\Omega_{m0}$ varies from 0.05 to 0.5, which we ignore, so the cosmology dependence is again only $D_L$. Completely ignoring the generalization introduces an error of <±2% for the low average redshift of our statistical sample, but we include it for completeness.

## 3.5. *Scatter*

Although the intrinsic scatter in mass at constant temperature about the M – T relation is expected to be low, it is not expected to be zero. But because the scatter is low, of all the parameters we discuss here its measurement is the most problematic. Fortunately, the effect of uncertainties in the scatter is to slide the cosmology constraints along their error ellipse. Thus while an incorrect measurement of the scatter propagates to an error on $\sigma_8$, the error ellipse changes less so. In addition, the effect is not large: Rasia et al. (2005) find that $\sigma_8$ decreases by about 5% when the assumed scatter doubles from 16% to 30%.

There are several reasons why this parameter is uncertain. First, we find $\sigma_{MT}/M = 0.128 \pm 0.087$ (68% confidence on one parameter) from the combined fit, which is only $1.5\sigma$ significant. Second, both empirical methods (Sections 3.1 and 3.2) only provide upper limits. Hence the measurement is most influenced by the simulations. Third, the scatter from the weak lensing mass method may be artificially increased due to the contribution of projected mass from large-scale structures, and the X-ray mass method may have artificially low scatter because such masses are proportional to the temperature at $r_{500}$. While these two effects tend to compensate each other, it would be unwise to assume they do so with any precision.

Stanek et al. (2006) point out that the scatter in the L – T relation contains information on the scatter in mass at constant temperature and at constant luminosity if the covariance between the two scatters is known. Unfortunately the scatter in mass at constant L or T and their covariance must be known to derive the scatter in mass at constant T or L, so this method is not a viable option at present. Stanek et al. (2006) do find $\sigma_{MT}/M = 0.19$ from an ensemble of 68 cluster simulations, which agrees with that from the simulations of Kravtsov et al. (2006, $0.20 \pm 0.05$) and our adopted value ($0.13 \pm 0.09$).

## 3.6. *Status of the Mass - Temperature Calibration*

The agreement among the three methods described in Sections 3.1, 3.2 and 3.3 is a significant advance and is probably no accident. The simulations are more realistic, include objects with the same masses as the data sample and they have been observed and analyzed the same way that actual data are. The biggest advance toward agreement between observations and simulations is not using an emission-weighted temperature, but rather folding the simulated spectra through the response of an actual detector (Mazzotta et al. 2004). The main deficiency of the simulations we used is the high fraction of baryons in stars (~40% at z = 0, see Nagai, Kravtsov & Vikhlinin, 2007) compared to observations. This cold gas may generate more non-thermal pressure support in the simulations than in actual clusters because, as it moves through the cluster, it can drive bulk motions into the hot gas.

The observations are also more carefully made and analyzed. All temperatures for the sample and M – T and L – T calibrations are from the same source or put onto that scale. The cluster masses are not extrapolated beyond the range of the observations for either the weak lensing or X-ray masses. X-ray masses are only for relaxed clusters using actual temperature and surface brightness profiles (i.e. all clusters are not assumed to be isothermal beta models in hydrostatic equilibrium), which are then scaled to apply to all clusters. The weak lensing observations use multi-color, wide-field data to derive aperture masses at large radii. This approach minimizes the mass sheet degeneracy, contamination by cluster galaxies, cluster centroiding errors and the effects of substructure.

Despite these advances some issues remain. Our temperature limit and M – T calibration imply a mass limit of h $M_{500,15}$ > 0.2 or h $M_{200,15}$ > 0.3 for a NFW mass profile with c = 5. Nord et al. (2008) show that calibrating the M – L relation above this mass requires using a flux limited sample with limit ~$10^{-13}$ erg $cm^{-2}$ $s^{-1}$ in order to avoid biasing (brightening) the derived M – L relation. At a redshift of 0.1 (all but 2 of our 48 clusters are within that redshift) this flux corresponds to a bolometric luminosity of ~3.1 x $10^{42}$ or a temperature of ~0.9 keV. There are no cluster surveys with flux or temperature limits of $10^{-13}$ erg $cm^{-2}$ $s^{-1}$ or 0.9 keV. Does this mean our M – T relation is heated relative to the mass-limited relation? It is difficult to apply this calculation to our case because the two empirical calibrations are not based on flux or temperature limited samples. However, since the biasing is ~ $\sigma_{MT}$/M, it will be about 5 times smaller than the corresponding bias for the M – L relation or ~10%. Such a bias, if it exists, is comparable to the statistical error on $A_{MT}$, which we marginalize over in any case.

We have attempted to assess how robust the adopted M – T calibration is by comparing it with that obtained with other samples. The closest weak lensing work to ours is Bardeau et al. (2007). It is indeed very close as all but one of the eleven objects in their sample (A1835) is in Hoekstra (2007) and the observations are the same, coming from the CFHT archive. Since the temperatures always come from Horner (2001), a comparison will only compare the cluster masses derived from the different methods. Bardeau et al. only report $M_{200}$. The error-weighted mean ratio of Hoekstra (with the Mahdavi revisions) to Bardeau $M_{200}$s is 0.93 ± 0.12, clearly good agreement.

Next, we examine the calibration coming from X-ray determined masses. Arnaud, Pointecouteau, & Pratt (2005) observed a sample of 10 relaxed clusters with XMM-Newton, 7 of which have Horner (2001) temperatures. Two of these 7 are also in the Vikhlinin et al. (2006) sample. Using the same procedure as in Section 3.2 we derive the M – T parameters from the Arnaud et al. (2005) sample shown in Figure 6. There is good agreement with the previously described results, although they are not completely independent. A few days after this paper was submitted, Vikhlinin et al. (2008) presented masses for a sample of 17 clusters, that is 7 new masses added to the sample analyzed in section 3.2. Twelve of these 17 clusters have Horner temperatures and the ratio of the Chandra to Horner temperatures remains 1.081 ± 0.008. However we deemed this sample large enough not to require converting the Chandra temperatures to the Horner scale. The M – T relation from the 12 Horner temperatures and Chandra masses agrees with that found in section 3.2, when derived as described therein. Vikhlnin et al. (2008) also derive a M – T relation, which has the same power law index as our joint fit, but our normalization is a factor 1.35 higher. This discrepancy comes almost entirely from their not including a correction for non-thermal pressure support, preferring to book keep this correction as a systematic error.

Finally, Rasia et al. (2005) present simulations of 99 clusters. These simulations are similar to those of Kravtsov et al. (2006) but use independent algorithms and codes. A comparison of the M – T relations from the two may give an indication how robust the simulation results are. The main differences appear to be the inclusion of metals by Kravtsov et al. but not by Rasia et al. and the amount of heating by feedback. Rasia et al have ~20% of the baryons in stars versus ~40% for Kravtsov et al. Rasia et al. (2005) give cluster spectroscopic-like temperatures, which Mazzotta et al. (2004) defined to reproduce the temperature measured by *Chandra*, *XMM-Newton* or ASCA detectors for clusters hotter than a few keV. We assume these spectroscopic-like temperatures are the same as Horner's ASCA temperatures as no central region is excluded. The M – T relation parameters are shown in Figure 6. There is mild disagreement (~1.9σ) with the fit to the Kravtsov et al. simulations described in Section 3.3.

Zhang et al. (2008) discuss some of these issues. Their sample and analysis procedure differs from ours in some important ways. The sample is at a higher redshift than the X-ray mass calibrators, but similar to the weak lensing calibrators. In fact part of their weak lensing sample is the Bardeau et al. (2007) sample, which is virtually identical to ours as described above. Zhang et al. (2008) derive X-ray hydrostatic masses for all objects in their sample, despite recognizing that about half of them exhibt signs of an unrelaxed merger. They do derive the masses using spatially resolved temperatures from XMM observations. The normalization of the M – T relation of Zhang et al. (2008) is identical with that from Arnaud et al. (2005) and Vikhlinin et al. (2006) (factor of 1.01 ± 0.07 difference) when the slopes are fixed to those found by the latter two works. Zhang et al. (2008) find the average ratio of weak lensing to X-ray masses for the Bardeau et al. (2007) sample for which they have X-ray data is 1.1 ± 0.2. This result confirms Mahdavi et al. (2008) (1.28 ± 0.15) for essentially the same sample but using different X-ray data (XMM vs. Chandra, respectively). The best agreement between the average ratio of lensing to X-ray masses comes when comparing a combined Bardeau et al. (2007) plus Dahle (2006) lensing sample with the X-ray masses of Zhang et al. (2008) when both masses are measured to the same radius, $r_{500}$ determined from X-ray observables. The average ratio is 1.09 ± 0.08. However the Dahle (2006) masses must be extrapolated to $r_{500}$, something we do not do. It is difficult to compare our two analyses, given that Zhang et al. do not adhere to some of our criteria (only use relaxed-looking clusters for X-ray masses, do not extrapolate any masses). It is probably safe to say that there are no strong disagreements and Zhang et al. find at least some evidence for non-hydrostatic support at the 10 – 20% level although possibly lower than what we use (39 ± 8%).

A summary statement on the status of the M – T calibration is the usual one: We are moving in the right direction; more can and should be done. Larger lensing and X-ray samples suitable for calibration are possible using extant data. Uniform analyses of a large fraction of the *Chandra* (Maughan et al., 2008) and *XMM-Newton* (Snowden et al., 2008) cluster databases, which can replace the Horner (2001) ASCA compilation, have appeared. Efforts need to be made to try to reduce the M – T relation intrinsic scatter. The higher spatial resolution *Chandra* and *XMM-Newton* data may help in this area by, for example, allowing the exclusion of the cluster centers. These data will permit using observables that may have even smaller scatter than temperature, for example $Y_X = kT\, M_{gas}$ (Kravtsov et al., 2006), Efforts to make more realistic simulations and analyze them as observers do will continue. The overall goal will be to find the systematic floor on the calibration. At present there is no strong evidence that the floor has been reached since the scatter in Figure 6 could be just statistical i.e. not due to systematics.

## 4. MASS AND TEMPERATURE FUNCTIONS

In this section we derive the temperature function for comparison to our cluster temperature observations. This calculation begins with the theoretically provided mass function, which is

$$n(M,z) = \frac{\rho_{b0}}{M} \frac{d\nu}{dM} f(\nu). \qquad (7)$$

Where $\rho_{b0}$ is the present background matter density, $\nu = 1.686 / \sigma(M,z)$, $\sigma(M,z) = \sigma(M) D(z)/D(0)$ is the rms mass fluctuation of the density field on scale M and D(z) is the growth factor to redshift z (equation A18 in Henry, 2000) and

$$\sigma^2(M) = \sigma_8^2 \frac{\int_0^\infty dk\, k^2\, P(k) \left[\frac{j_1(kR)}{kR}\right]^2}{\int_0^\infty dk\, k^2\, P(k) \left[\frac{j_1(k8h^{-1})}{k8h^{-1}}\right]^2}. \qquad (8)$$

Here R = 9.5 h$^{-1}$ Mpc (h $M_{15}/\Omega_{mo}$)$^{1/3}$ where $M_{15}$ is the mass in units of $10^{15}$ M☉, k is the spatial wavenumber, $P(k) \sim k^{n_s} T^2(k)$ and T(k) the transfer function from Eisenstein & Hu (1998), which is a function of h, $\Omega_{m0}$, $\Omega_{b0}$ and the present CMB temperature (taken to be 2.728 K). This notation is standard, but note the subtle difference between the cluster temperature, kT, and the transfer function, T(k). We take the values of h, $n_s$ and $\Omega_{b0}$ their variance and covariance in the case of the latter two parameters from the five year WMAP results (Dunkley et al. 2008).

A number of forms for f(ν) have been proposed (e.g. Press & Schechter 1974; Sheth & Torman 1999; Warren et al. 2006; Tinker et al. 2008). We use the Jenkins et al. (2001) form of the mass function for which

$$f(\nu) = \frac{A_{MF}}{\nu} \exp\left(-\left|\ln\left(\frac{\nu}{1.686}\right) + B\right|^\varepsilon\right). \qquad (9)$$

We obtain $A_{MF}$, B and ε by fitting this form to the Hubble volume simulation local mass function for $M_{500}$ given in Figure 20 of Evrard et al. (2002). We used the ΛCDM model simulation, which had h = 0.7, $\Omega_{mo} = 0.3$, $\Omega_{bo} h^2 = 0.0196$, $\sigma_8 = 0.9$, $n_s = 1$. By design the masses use the same definition as the M – T calibration presented in Section 3 and the cosmology is very similar to what we find or marginalize over here. Table 1 gives the best fit values of $A_{MF}$, B and ε and information on the covariance matrix of the later two. Figure 7 shows the error ellipse for B and ε, while Figure 8 compares the best fit with the data. We include a 20% systematic uncertainty on $A_{MF}$, based on recent estimates of the effects of baryonic physics that can change the halo mass by about ±10% (Stanek, Rudd, & Evrard, in preparation) and by comparison of our mass function to that of Tinker et al. (2008). Figure 9 shows the comparison for $\Omega_{m0} = 0.24$ and 0.30 and overdensity with respect to the mean (used by Tinker et al. instead of with respect to critical that we use) of $500/\Omega_{m0}$. The Tinker function agrees with the Jenkins function at all masses for $\Omega_{m0} = 0.24$, but lies above it at all masses for $\Omega_{m0} = 0.30$. A conservative statement is that both functions may have a normalization error of ±20%.

The differential temperature function comes from the mass function via the chain rule

$$n(kT) = \alpha_{MT} \frac{\rho_{b0}}{kT} \frac{d\nu}{dM} f(\nu). \qquad (10)$$

Now

$$R = 9.5h^{-1}\text{Mpc}\left[\frac{A_{MT}}{\Omega_{m0}}(kT)^{\alpha_{MT}}\frac{1}{E(z)}\frac{D_L(1,\Omega_{m0},z)}{D_L(1,0.3,z)}\right]^{1/3}. \qquad (11)$$

It is more convenient to express the derivative as $-\nu/2\ d\ln\sigma^2(M)/dM$. Then after some algebra we find

$$n(kT,z) = 9.25\times 10^{-5}(h^{-1}\text{Mpc})^{-3}\text{keV}^{-1}\Omega_{m0}\frac{\alpha_{MT}E(z)D_L(1,0.3,z)}{A_{MT}D_L(1,\Omega_{m0},z)}(kT)^{-(\alpha_{MT}+1)}$$

$$\frac{\int_0^\infty dy\ y^{(n_s+2)}\ T^2(y/R)\ j_1(y)j_2(y)/y}{\int_0^\infty dy\ y^{(n_s+2)}\ T^2(y/R)\left[j_1(y)/y\right]^2}$$

$$A_{MF}\exp\left(-\left|\ln\left(\frac{D(0)}{\sigma_8 D(z)}\left\{\left(\frac{R}{8h^{-1}}\right)^{n_s+3}\frac{\int_0^\infty dy\ y^{(n_s+2)}\ T^2(y/8h^{-1})\left[j_1(y)/y\right]^2}{\int_0^\infty dy\ y^{(n_s+2)}\ T^2(y/R)\left[j_1(y)/y\right]^2}\right\}^{1/2}\right)+B\right|^\varepsilon\right), \qquad (12)$$

where $j_2$ is the Spherical Bessel function of the first kind, order two.

## 5. RESULTS

We perform a maximum likelihood fit of equation (12) to 48 cluster (kT, z) pairs, marginalizing over 12 "uninteresting" parameters, to derive $\sigma_8$ and $\Omega_{m0}$. We list these uninteresting parameters in Table 1. That is, we marginalize over *all* systematic uncertainties of which we are aware. The uninteresting parameters are the three mass function and six cluster physics (M – T and L – T) parameters and three cosmological parameters (h, $n_s$ and $\Omega_{b0}h^2$). As we will see, the cosmological parameters have virtually no effect on the best fit or error on $\sigma_8$ and $\Omega_{m0}$. The effective number of marginalized parameters is reduced to eight from covariances among them. The negative of the natural logarithm of the likelihood function is equation (7) of H04:

$$S = -2\sum_{i=1}^{N}\ln\left[\int dkT\ n(kT,z_i)\ \frac{\exp\left[-(kT_i-kT)^2/2\sigma_i^2(kT)\right]}{\sqrt{2\pi\sigma_i^2(kT)}}\Omega(kT,z_i)\frac{d^2V}{d\Omega dz}\right]$$

$$+2\int_{kT\min}^{kT\max}dkT\int_{z\min}^{z\max}dz\ n(kT,z)\ \Omega(kT,z)\frac{d^2V}{dzd\Omega}+\Delta S(\text{uninteresting}) \qquad (13)$$

where $\Delta S$(uninteresting) is the increase in likelihood as all the parameters in Table 1 but $\Omega_{m0}$ and $\sigma_8$ deviate from their best fit values. This increase in likelihood comes from the fits in Sections 2, 3 or 4 or from WMAP5. The Gaussian dispersion includes contributions from the individual cluster temperature errors and scatter in the M – T relation summed in quadrature: $\sigma_i^2 = \sigma_{kTi}^2 + (kT_i\ \sigma_{MT}/M/\alpha_{MT})^2$. The integral of the Gaussian distribution is absent from the second term of S because it is possible to perform it analytically. The best estimates of the model parameters are obtained by minimizing S. Confidence regions for the best estimates are obtained by noting that S is distributed as $\chi^2$ with the number of degrees of freedom equal to the number of interesting parameters. The values of $z_{min}$, $z_{maz}$, $kT_{min}$ and $kT_{max}$ are 0.00, 0.20, 3 and 12 keV respectively.

We present in Figure 10 the 68% and 95% confidence contours in the $\sigma_8$ and $\Omega_{m0}$ plane. Figure 10a shows the statistical uncertainties only, while Figures 10b, 10c and 10d show the effect of marginalizing over increasingly more systematic uncertainties. Marginalizing over cluster physics (L – T and M – T) lengthens the error ellipse; marginalizing over the mass function widens the error ellipse and marginalizing over cosmology has almost no effect (to our knowledge first pointed out by Voevodkin & Vikhlinin, 2004). The error ellipse is described by $\sigma_8 (\Omega_{m0}/0.32)^{\alpha} = 0.86 \pm 0.04$ (68% confidence for 2 parameters) with $\alpha = 0.30$ (0.41) for $\Omega_{m0} \leq 0.32$ ($\Omega_{m0} \geq 0.32$). Figure 11 compares the best fit temperature function with the data from our HIFLUGCS sample. The fit is obviously very good.

Table 1 gives the error on $\sigma_8$ and $\Omega_{m0}$ from propagating the errors of all 12 uninteresting parameters individually. The largest contributor comes from the M – T relation. We also give the increase in likelihood as the 12 uninteresting parameters assume their 68% confidence values. When this parameter is near 1, then the sample of 48 does not constrain that particular excursion.

### 5.1. *Cosmic Variance*

Virtually the entire extragalactic sky was surveyed to a uniform flux level to find the clusters in our sample. Thus, this sample is about the only one that will be available with its selection criteria. How representative is it? Or put another way, what is the sample or cosmic variance on the best fitting parameters from not being able to average over more samples? We answer this question by calculating the Fisher matrix, including the effects of cosmic variance, for our sample. The Fisher matrix is the inverse of the covariance matrix of parameters $p = (\Omega_{m0}, \sigma_8)$. We perform an approximate calculation for the cosmic variance part of the Fisher matrix, since, for our survey, it is expected to be small compared to the Poisson part (Hu & Kravtsov, 2003). We begin by dividing the observation space into 10 redshift and 100 temperature pixels. The redshift pixels have constant linear width $\Delta z = 0.02$ and the temperature pixels have constant logarithmic width $\Delta \log kT = \log(4)/100$ (or $\Delta kT = kT \ln(4)/100$). We remove from this grid the 81 low – temperature, high – redshift pixels that have less then $10^{-6}$ expected clusters in them, leaving 919 pixels. This removal is to increase the numerical stability of the calculation. We verified that increasing the number of pixels to 20 in redshift and 300 in log kT gives the same Fisher matrix for the pure Poisson case. The formalism comes from section 6 of Chapter XIII of the Dark Energy Task Force Final Report (Albrecht et al., 2006). The Fisher matrix is

$$F_{\mu\nu} = \sum_{ij} \frac{\partial \overline{N}_i}{\partial p_\mu} \left(C^{-1}\right)_{ij} \frac{\partial \overline{N}_j}{\partial p_\nu}. \qquad (14)$$

Here 
$$\overline{N}_i = \Delta z_i \, \Delta kT_i \, \frac{d^2V}{d\Omega dz} \int dkT \, n(kT, z_i) \, \frac{\exp\left[-(kT_i - kT)^2 / 2\sigma^2\right]}{\sqrt{2\pi\sigma^2}} \Omega(kT, z_i) \qquad (15)$$

is the expected number of clusters in the $i^{th}$ (z, kT) pixel (i = 1, 919) and $\sigma = kT \, \sigma_{MT}/M/\alpha_{MT}$. We use equation (2) from Levine, Schultz & White (2002) for the numerical derivative:

$$\frac{\partial \overline{N}_i}{\partial p_\mu} = \frac{\overline{N}_i(p_\mu \times \Delta\theta) - \overline{N}_i(p_\mu / \Delta\theta)}{2p_\mu \ln(\Delta\theta)} \tag{16}$$

with $\Delta\theta = 1.0001$. The matrix $C_{ij} = \overline{N}_i \delta_{ij} + S_{ij}$, where the first term is the Poisson variance and the sample variance is

$$S_{ij} = b_i \overline{N}_i b_j \overline{N}_j \int \frac{d^3k}{(2\pi)^3} W_i^*(k) W_j(k) P(k). \tag{17}$$

$W_i(k)$ is the Fourier transform of the pixel window normalized such that $\int d^3x\, W_i(x) = 1$ and the average bias of the selected clusters is

$$b_i = \frac{\Delta z_i \Delta kT_i}{\overline{N}_i} \frac{d^2 V}{d\Omega dz} \int dkT\, n(kT, z_i)\, b(kT, z_i)\, \frac{\exp\left[-(kT_i - kT)^2 / 2\sigma^2\right]}{\sqrt{2\pi\sigma^2}} \Omega(kT, z_i). \tag{18}$$

The bias is

$$b(kT, z) = 1 + \frac{a\delta_c^2 / \sigma^2(kT, z) - 1}{\delta_c} + \frac{2p}{\delta_c \left[1 + (a\delta_c^2 / \sigma^2(kT, z))^p\right]} \tag{19}$$

with $a = 0.75$, $p = 0.3$ and $\delta_c = 1.686$.

The approximation we make is to assume that our sample comes from the whole sky, not just the 8.14 steradians of the HIFLUGCS. In this case, the normalized redshift pixel is a shell at proper distance $R \pm \delta R$

$$W(r) = \frac{3(H(R + \delta R - r) - H(R - \delta R - r))}{4\pi((R + \delta R)^3 - (R - \delta R)^3)}, \tag{20}$$

with $R = D_L/(1 + z)$ and $H$ is the unit step function. The Fourier transform is

$$W(k) = \frac{3(R + \delta R)^3}{(R + \delta R)^3 - (R - \delta R)^3} \frac{j_1(k(R + \delta R))}{k(R + \delta R)} - \frac{3(R - \delta R)^3}{(R + \delta R)^3 - (R - \delta R)^3} \frac{j_1(k(R - \delta R))}{k(R - \delta R)}. \tag{21}$$

We show in Figure 12 the calculated Poisson and Poisson + cosmic variance error ellipses and the actual Poisson error ellipse. There are two things to note. The calculated and actual Poisson ellipses agree very well. This is the first published such comparison of which we are aware. The cosmic variance is a small increase over the purely Poisson errors. The calculated values of the cosmic variance errors are: $\Delta\Omega_{m0} = 0.040$ and $\Delta\sigma_8 = 0.026$, and we include these errors in Table 1. We find $\Delta\sigma_8/\sigma_8 = 3.1\%$, while Evrard et al. (2002) find 3.9% by analyzing local cluster temperature samples comprised of 30 objects on average (vs. 48 for our sample) from the Hubble Volume numerical simulations.

## 6. COMPARISON WITH OTHER RESULTS

A good way to assess whether there are undetected systematic effects in a measurement is to compare it to other independent measurements of the same thing. If all measurements agree within the errors, then the errors are likely to be correctly estimated. An often-discussed example is the Hubble parameter, which has ranged from 465 ± 50 km s$^{-1}$ Mpc$^{-1}$ (Hubble, 1929) to 95 ±10 km s$^{-1}$ Mpc$^{-1}$ (de Vaucouleurs, 1982) to 50.3 ± 4.3 km s$^{-1}$ Mpc$^{-1}$ (Sandage & Tammann, 1976). At least two of these results had systematic errors not reflected in the quoted errors. Thus in this section we compare our measurements of $\Omega_{m0}$ and $\sigma_8$ with the most recent independent measurements made with different techniques.

### 6.1 *WMAP5*

Figure 10d shows that the cluster results agree with those of the five-year WMAP analysis at the 68% confidence level. We estimate the joint constraints by fitting a two-dimensional Gaussian likelihood to the WMAP5 contours, then adding it to the X-ray cluster likelihood. We show the resulting joint likelihood in Figure 13, which is characterized as $\Omega_{m0} = 0.30^{+0.03}_{-0.02}$ and $\sigma_8 = 0.85^{+0.04}_{-0.02}$ (68% confidence for two parameters). Komatsu et al. (2008) combined WMAP5 with baryon acoustic oscillations + supernovae constraints and derived consistent results with similar sized errors: $\Omega_{m0} = 0.279 \pm 0.023$ and $\sigma_8 = 0.817 \pm 0.041$ (68% confidence for two parameters).

### 6.2 *Other Cluster Results*

#### 6.2.1 *X-ray Selection*

We compare in Figure 14 the most recent cosmological constraints from X-ray selected clusters using X-ray temperatures of 48 objects with z ≤ 0.2 (our work here), galaxy velocities of 72 objects with z ≤ 0.1 (Rines, Diaferio & Natarajan, 2007; 2008), X-ray luminosities of 242 objects with z < 0.7 (Mantz et al., 2008) and $Y_X$ of 49 objects with 0.025 ≤ z ≤ 0.22 (Vikhlinin, 2008 private communication). Rines et al. (2007) have 7 uninteresting parameters, all fixed. Mantz et al. (2008) have 14 uninteresting parameters in their analysis, of which they fit 2, fixed 4 and marginalized 8. Vikhlinin (2008) gives the largest systematic effect for their measurement, which comes from the absolute mass calibration. We include these systematic effects as estimated by each work in the errors shown in Figure 14. The agreement among these four measurements is less than or about equal to the 1σ estimated systematic errors.

Comparing our results with those of Vikhlinin (2008) provides a good indication of the systematic errors of each, since we do an independent analysis of independent data for nearly the same objects (39 of our 48 objects are in his sample). Although Vikhlinin uses the Tinker et al. (2008) mass function compared to our use of the Jenkins et al. (2001) form, the largest difference between our work is the normalization of the M – T relation as can be seen from the error propagation in Table 1. Our constraints are disjoint with the optimistic 68% confidence, 1 parameter purely statistical errors, but overlap with the same errors if we use Vikhlinin's M – T relation in our analysis. However, as shown in Figure 14, our contours nearly touch each other after including systematic errors, indicating the systematics have been included correctly.

X-ray selected samples now provide consistent constraints independent of the observable. Evrard et al. (2008) have reached the same conclusion for temperature and galaxy velocity dispersion observables, finding $\sigma_8 = 0.88$ for $\Omega_{m0} = 0.32$ vs. our $0.86 \pm 0.04$.

### 6.2.2 *Optical Selection*

Optical surveys are another way to build cluster catalogs and there are many reported cosmological constraints using such samples, see Figure 9 of Rines, Diaferio & Natarajan (2007) for some examples. We compare our results with two surveys that use richness as the observable. The survey and analysis closest to ours is the Red-sequence Cluster Survey (Gladders et al., 2007, RCS) of 956 richness-selected clusters with $0.3 < z < 0.95$ found in 72.1 square degrees. There are 10 uninteresting parameters associated with this analysis, of which Gladders et al. (2007) fit 5 and fixed 5. We compare their results with ours in Figure 15. The agreement is $\sim 1.3\sigma$ for $\Omega_{m0} = 0.25$. A somewhat similar analysis of 13,823 maxBCG clusters in 7500 square degrees of the Sloan Digital Sky Survey with $0.1 < z < 0.3$ is presented by Rozo et al. (2007). The results are not directly comparable to ours because Rozo et al. impose an effective prior of $\Omega_{m0} = 0.24 \pm 0.04$ from cosmic microwave background and supernovae results. Nevertheless, their $\sigma_8 = 0.92 \pm 0.10$ (1 $\sigma$) agrees with ours in the relevant $\Omega_{m0}$ range, as we show in Figure 15.

### 6.3 *Cosmic Shear*

Large-scale mass structures between a distant source and an observer induce a weak gravitational lensing shear in the shape of the distant source. This signal yields a direct measure of the projected matter power spectrum, i.e. a measure of $\sigma_8$ in a very complementary way to the X-ray cluster technique. In Figure 16 we compare the constraints from this paper with those from the 100 Square Degree Weak Lensing Survey, which has the largest solid angle of any such survey so far (Benjamin et al., 2007). The figure shows that constraints from the two methods agree at the $\sim 68\%$ confidence level. Other recent shear results are consistent with the 100 Square Degree Survey (e.g. Jarvis et al., 2006).

We may summarize this section as follows: nearly all of the most recent results from a variety of independent techniques are consistent with each other and with this paper.

### 7. CONCLUSIONS

The most important conclusion of this paper is that we find no evidence of additional systematic errors beyond what we have considered for the X-ray cluster temperature function measurement of $\sigma_8$. We base these conclusions on agreement with the most recent WMAP and weak lensing cosmic shear measurements within that error. Also significant is the most recent constraints on $\sigma_8$ and $\Omega_{m0}$ from X-ray selected cluster samples are independent of observable for X-ray luminosity, temperature or galaxy velocities. The cluster temperature results presented here are succinctly summarized as $\sigma_8 (\Omega_{m0}/0.32)^\alpha = 0.86 \pm 0.04$ with $\alpha = 0.30$ (0.41) for $\Omega_{m0} \leq 0.32$ ($\Omega_{m0} \geq 0.32$). The joint cluster temperature + WMAP5 constraints are $\Omega_{m0} = 0.30^{+0.03}_{-0.02}$ and $\sigma_8 = 0.85^{+0.04}_{-0.02}$ (all at 68% confidence for two parameters).

We thank Elena Rasia and Alexey Vikhlinin for kindly providing results from their work in digital form and for informative discussions. This work was supported by NASA grants NNX07AE73G and NNX07AP01G (JPH), NSF grant AST-0708150 (AEE) and NSERC Discovery Grants (AB, HH).

APPENDIX A

In this paper we fit the luminosity – temperature and mass – temperature relations as well as the mass function to a Jenkins form. We give some details of those fits here.

We use the BCES bisector linear least squares fit of Akritas & Bershady (1996) when fitting the L – T and M – T relations. This method allows for errors on both variables that may be correlated and different for each data point plus intrinsic scatter in the data beyond the errors. We fit a power law relation between the two variables, $y = A\, x^{\alpha}$, and linearize it in the usual way by taking the logarithm. The publicly available BCES software computes the variances of the fit parameters. We also calculate their covariance according to equation 31 of Akritas & Bershady (1996). For purposes of visualization, we calculate the error ellipse from the variances and covariance as follows. The semimajor and semiminor axes of the ellipse are

$$\sqrt{\Delta\chi^2_\nu}\sqrt{\frac{\sigma_x^2 + \sigma_y^2 \pm \sqrt{(\sigma_x^2 - \sigma_y^2)^2 + 4\rho\sigma_x\sigma_y}}{2}}$$

where $\sigma_x^2$, $\sigma_y^2$, $\rho\sigma_x\sigma_y$ and $\Delta\chi^2_\nu$ are the variance of x, variance of y, covariance of x and y and increase in chi squared for a given confidence level and number of parameters, respectively. The symmetric covariance matrix is comprised of the first three quantities, thus it may be derived from the information in Table 1. Note that $\rho$ is just the usual correlation coefficient. The quantity $\Delta\chi^2_\nu$ is 1, 2.3, 3.53 and 6.17 for 68% confidence for one, two or three parameters and 95% confidence for two parameters, respectively. The angle of the semimajor axis counterclockwise with respect to the x axis is

$$\alpha = \frac{1}{2}\tan^{-1}\left(\frac{2\rho\sigma_x\sigma_y}{\sigma_x^2 - \sigma_y^2}\right).$$

This equation has two solutions that correspond to the semimajor and semiminor axes. Care must be taken to select the desired one.

We estimate the intrinsic scatter $\delta y$ in y about the relation $y = a\, x + b$ using

$$\delta y^2 = \frac{1}{N-2}\sum_{i=1}^{N}\left[(y_i - ax_i - b)^2 - (\sigma_{yi}^2 + a^2\sigma_{xi}^2)\right],$$

where N is the number of data pairs.

Errors on the covariance and scatter come from a jackknife procedure. Let $t_N$ be a statistic calculated from a sample of size N. Form N subsamples of size N-1 by dropping a data pair in succession. The jackknife estimate of the variance of $t_N$ is

$$\sigma_{tN}^2 = \frac{N-1}{N} \sum_{i=1}^{N} (t_{N-1,i} - \bar{t}_{N-1})^2$$

(Lupton, 1993, page 46), where $\bar{t}_{N-1}$ is the average of t over the N subsamples.

The fit procedure for the mass function is standard chi squared with the following deviation. When fitting a multiplicative parameter times a function, A f(x), the best fit value of A may be determined analytically to be

$$\log(A) = \frac{\sum_{i=1}^{N} (\log(y_i / f(x_i))) / \sigma(\log y_i)^2}{\sum_{i=1}^{N} 1 / \sigma(\log y_i)^2}.$$

For data with equal errors, this equation simplifies to yield A is the geometrical mean of $y_i/f(x_i)$.

## APPENDIX B

Sometimes the L – T and M – T relations are parameterized differently than what we have done. The difference is the temperatures pivot about (are divide by) a different value than the 1 keV used by us. A pivot point of 5 keV is sometimes used. More generally, a pivot point change from 1 to a new pivot a is a coordinate transformation in the space of fitting parameters from p = (log A, α) to π = (log A + α log a, α). Under this transformation the covariance matrix goes from C to J C J$^T$ where

$$J_{i,j} = \frac{\partial \pi_i}{\partial p_j} = \begin{pmatrix} 1 & \log a \\ 0 & 1 \end{pmatrix}$$

The transformed covariance matrix is diagonal when a = dex(-ρ $\sigma_{\log A}/\sigma_\alpha$) The intrinsic dispersion and its error are unchanged by this transformation because the values from the fit are the same, only the functional form changed. From the covariance matrices derived from the data in Table 1 for a pivot of 1, we see that the L – T relation is diagonalized for a = 4.95 keV and the M – T relation for a = 3.15 keV. As a worked example, we refit the L – T and joint M – T relations for a pivot of 5 keV. That is we fit $L_{44}$(bol) = A′$_{LT}$ (kT/5)$^{\alpha_{LT}}$ and h $M_{500,15}$ = A′$_{MT}$ (kT/5)$^{\alpha_{MT}}$. We find: log(A′$_{LT}$) = 0.7166 ± 0.0360, $\alpha_{LT}$ = 3.0870 ± 0.2370, ρ = 0.0300; log(A′$_{MT}$) = -0.3835 ± 0.0197, $\alpha_{MT}$ = 1.5376 ± 0.0653, ρ = 0.6660. It is easy to verify that all quantities transform as expected. The transformed L – T relation has almost zero covariance between the parameters while the transformed M – T relation retains substantial covariance, both as expected.

Table 1 Parameters and Error Propagation

| Symbol | Type | Value | Range (ΔLike=1) | $\Delta\Omega_{m0}$ | $\Delta\sigma_8$ | $\Delta S^a$ |
|---|---|---|---|---|---|---|
| Cosmology | | | | | | |
| $h$ | Prior | 0.719 | 0.692 – 0.745 | 0.00 – 0.00 | 0.00 – 0.00 | 0.64 – 1.20 |
| $n_s{}^b$ | Prior | 0.963 | 0.948 – 0.977 | 0.00 – 0.00 | 0.00 – 0.00 | 1.00 – 1.00 |
| $\Omega_{b0}h^{2\,b}$ | Prior | 0.02273 | 0.02211 – 0.02335 | 0.00 – 0.00 | 0.00 – 0.00 | 1.00 – 1.00 |
| $\Omega_{m0}$ | Fit | | 0.10 (0.02) 0.90 | | | |
| $\sigma_8$ | Fit | | 0.55 (0.01) 1.00 | | | |
| Mass Function | | | | | | |
| $A_{MF}$ | Prior | 0.148 | 0.119 – 0.178 | 0.04 – -0.03 | -0.01 – 0.01 | 0.95 – 0.98 |
| $B^c$ | Prior | 0.829 | 0.827 – 0.830 | -0.01 – 0.00 | 0.01 – 0.00 | 0.98 – 1.02 |
| $\varepsilon^c$ | Prior | 3.96 | 3.98 – 3.94 | -0.01 – 0.00 | 0.01 – 0.00 | 0.98 – 1.02 |
| Cluster Physics | | | | | | |
| $\log(A_{MT})^d$ | Prior | -1.4582 | -1.4939 – -1.4225 | 0.00 – 0.09 | 0.00 – -0.08 | 0.00 – 0.91 |
| $\alpha_{MT}{}^d$ | Prior | 1.5376 | 1.6029 – 1.4723 | 0.00 – 0.09 | 0.00 – -0.08 | 0.00 – 0.91 |
| $\sigma_{MT}/M$ | Prior | 0.13 | 0.04 – 0.22 | -0.13 – 0.00 | 0.07 – 0.00 | 3.32 – 0.00 |
| $\log(A_{LT})^e$ | Prior | -1.4411 | -1.6096 – -1.2726 | 0.02 – 0.05 | -0.02 – 0.04 | 1.06 – 3.49 |
| $\alpha_{LT}{}^e$ | Prior | 3.0870 | 3.3240 – 2.8500 | 0.02 – 0.05 | -0.02 – 0.04 | 1.06 – 3.49 |
| $\sigma_{\log(L)T}$ | Prior | 0.252 | 0.220 – 0.284 | -0.01 – -0.03 | 0.01 – 0.02 | 0.73 – 9.63 |
| | | | | | | |
| Sample Variance | | | | -0.04 – 0.04 | -0.03 – 0.03 | |

Prior = Priors on the $\Omega_{m0}$ and $\sigma_8$ fits, they are also the uninteresting parameters
[a]Including ΔS(uninteresting), which is +1 here   [b]Correlated   [c]Correlated $\rho$ = -1.0000   [d]Correlated $\rho$ = -0.9116   [e]Correlated $\rho$ = -0.9769

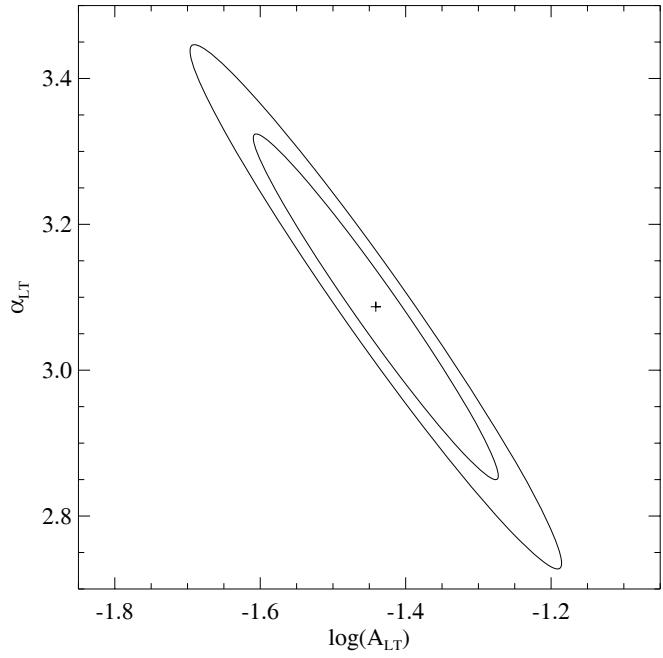

Fig. 1. Error ellipse for the luminosity – temperature relation parameters corresponding to 68% confidence for one and two parameters. A Hubble parameter of h = 0.7 was assumed to derive $A_{LT}$.

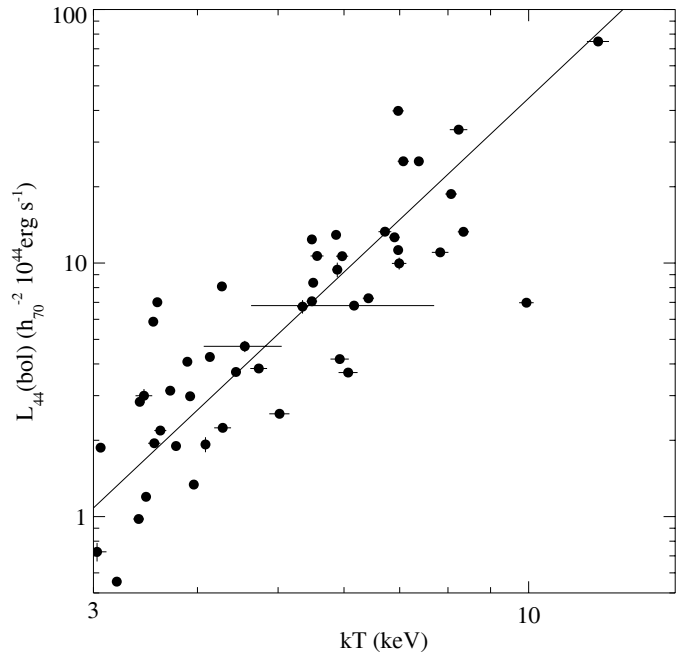

Fig. 2. Bolometric luminosity plotted as a function of Horner temperature. The best fitting luminosity – temperature relation is overlaid. The hottest cluster is A2163, which at z = 0.201 is just outside the z cut for the sample of 48 used for cosmology.

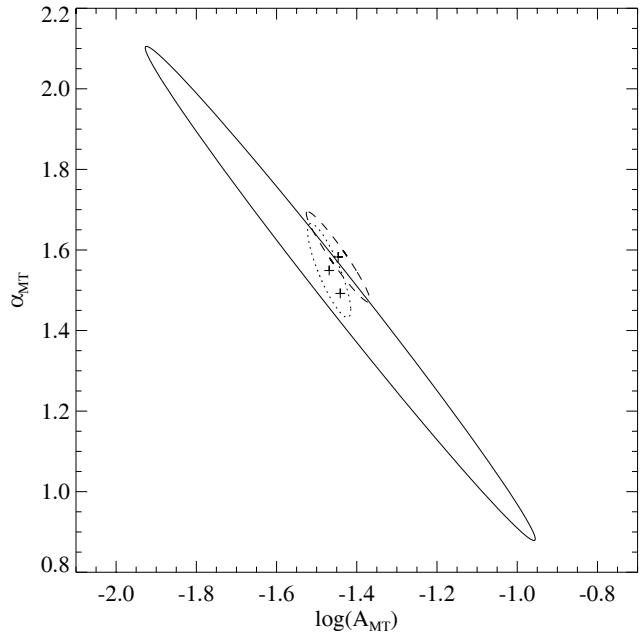

Fig. 3. Error ellipses for mass – temperature relation fits corresponding to 68% confidence for two parameters. The solid line is from the weak lensing masses and X-ray temperatures, the dashed line is from the X-ray hydrostatic equilibrium masses and X-ray temperatures and the dotted line is from the simulated masses and temperatures.

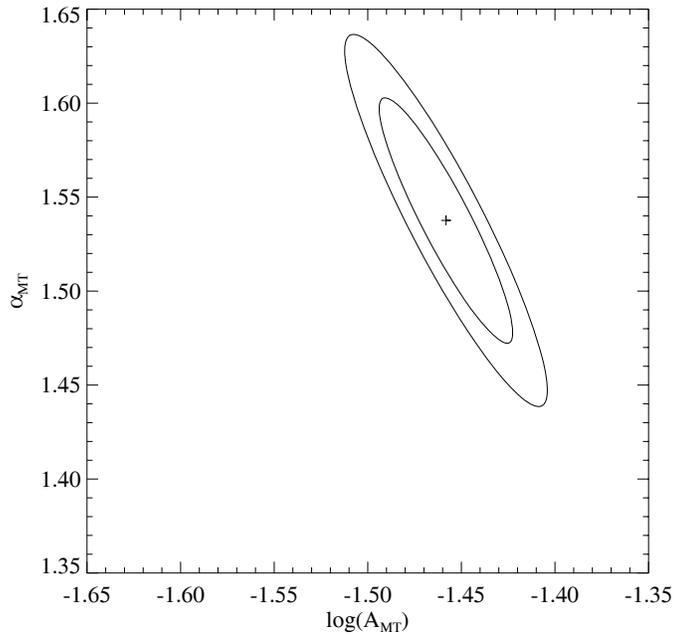

Fig. 4. Error ellipse for the mass– temperature relation parameters corresponding to 68% confidence for one and two parameters.

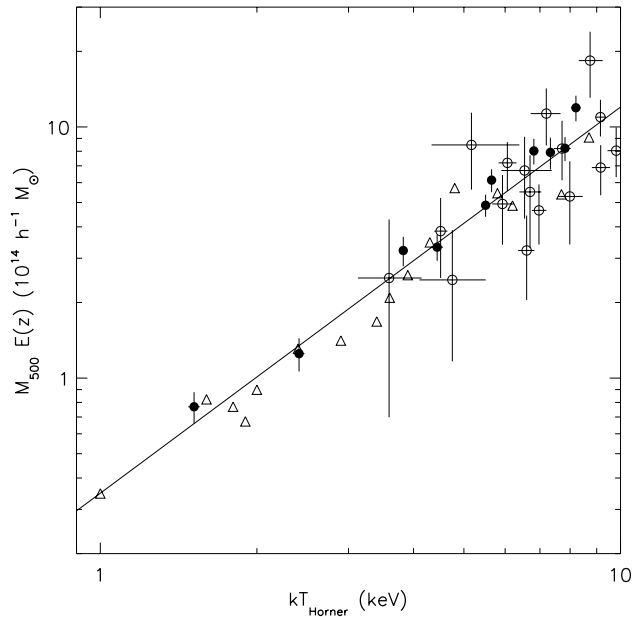

Fig. 5. Cluster mass plotted as a function of Horner temperature. The best joint fit mass– temperature relation is overlaid. Open circles are weak lensing masses, closed circles are X-ray hydrostatic equilibrium masses with the corrections described in the text and the open triangles are simulated masses and temperatures.

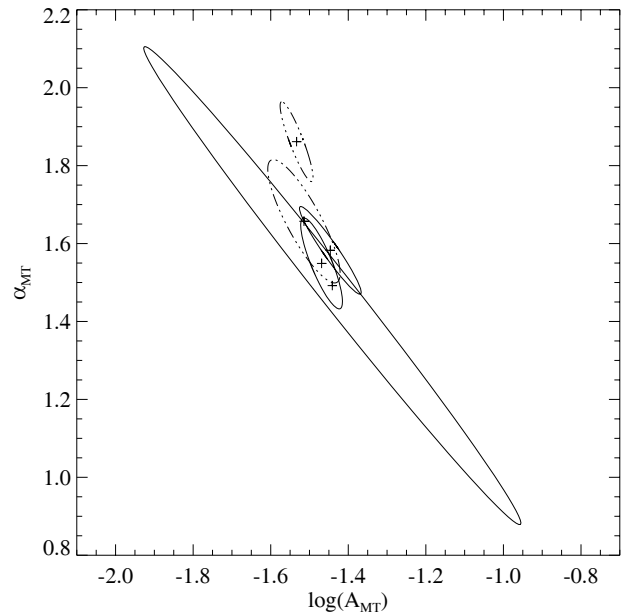

Fig. 6. Sixty-eight percent confidence error ellipses for two parameters from alternative determinations (triple dot dash) of the M − T relation compared to the determinations described in Sections 3.1, 3.2, and 3.3. The larger alternative ellipse is derived from the X-ray mass sample of Arnaud et al. (2005) and the smaller from the Rasia et al. (2005) simulations.

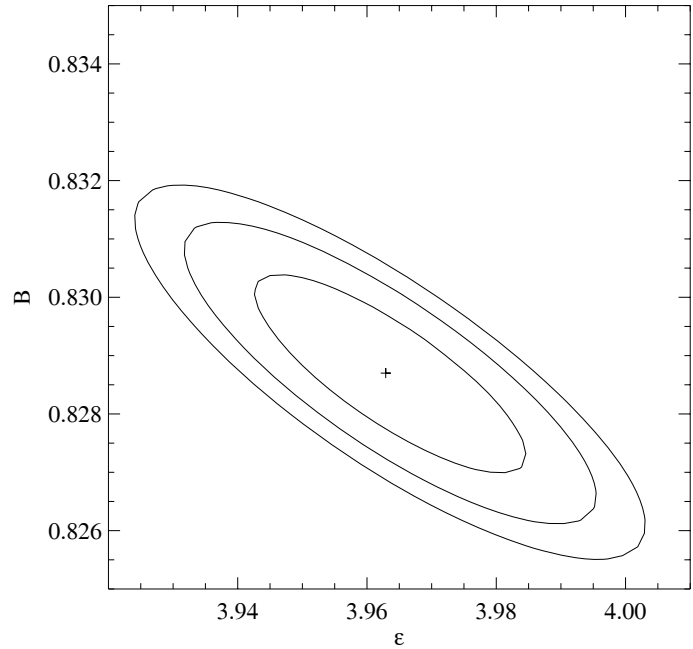

Fig. 7. Error ellipse for the Jenkins mass function parameters B and ε corresponding to 68% confidence for one, two and three parameters.

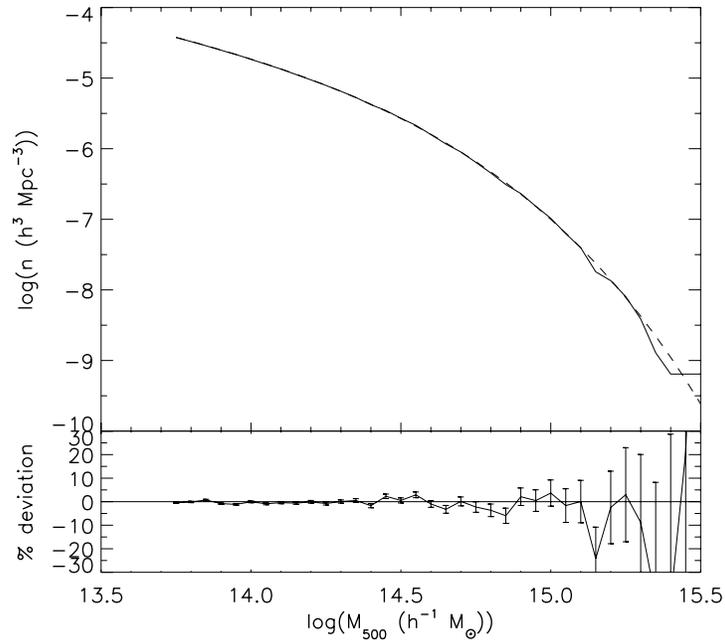

Fig. 8. (*Top*) Critical spherical overdensity of 500 (SO500c) mass function derived from the z = 0 Hubble Volume simulations (*solid line*) with best fitting Jenkins mass function overlaid (*dashed line*). (*Bottom*) Percent deviation in number density between the simulation data and the fit. Errors assume Poisson statistics in each mass bin.

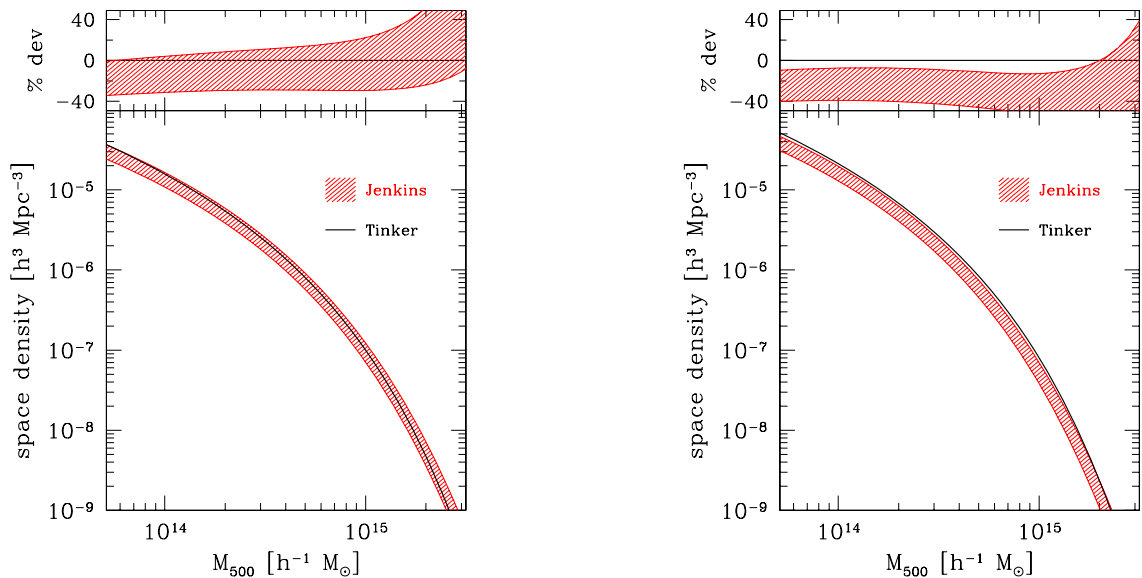

Fig. 9. (*Bottom*) Comparison of the SO500c z = 0 mass function of Jenkins used here to that of Tinker et al. (2008). The shaded region is the 68% confidence region for the Jenkins function. (*Top*) Percent deviation between Jenkins and Tinker functions. We fit over a temperature range corresponding to h $M_{500}$ = 0.19 – 1.16. a. (*Left*) $\Omega_{m0}$ = 0.24 b. (*Right*) $\Omega_{m0}$ = 0.30

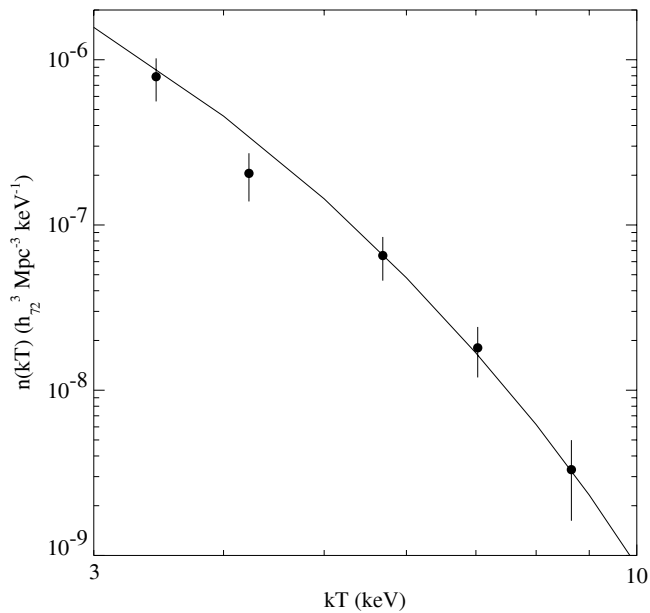

Fig. 11. The cluster temperature function derived from our HIFLUGCS sample compared to the best fitting function. The data are binned into five equal logarithmic temperature bins from 3 – 10 keV and are plotted at the average temperature of the clusters in the bin.

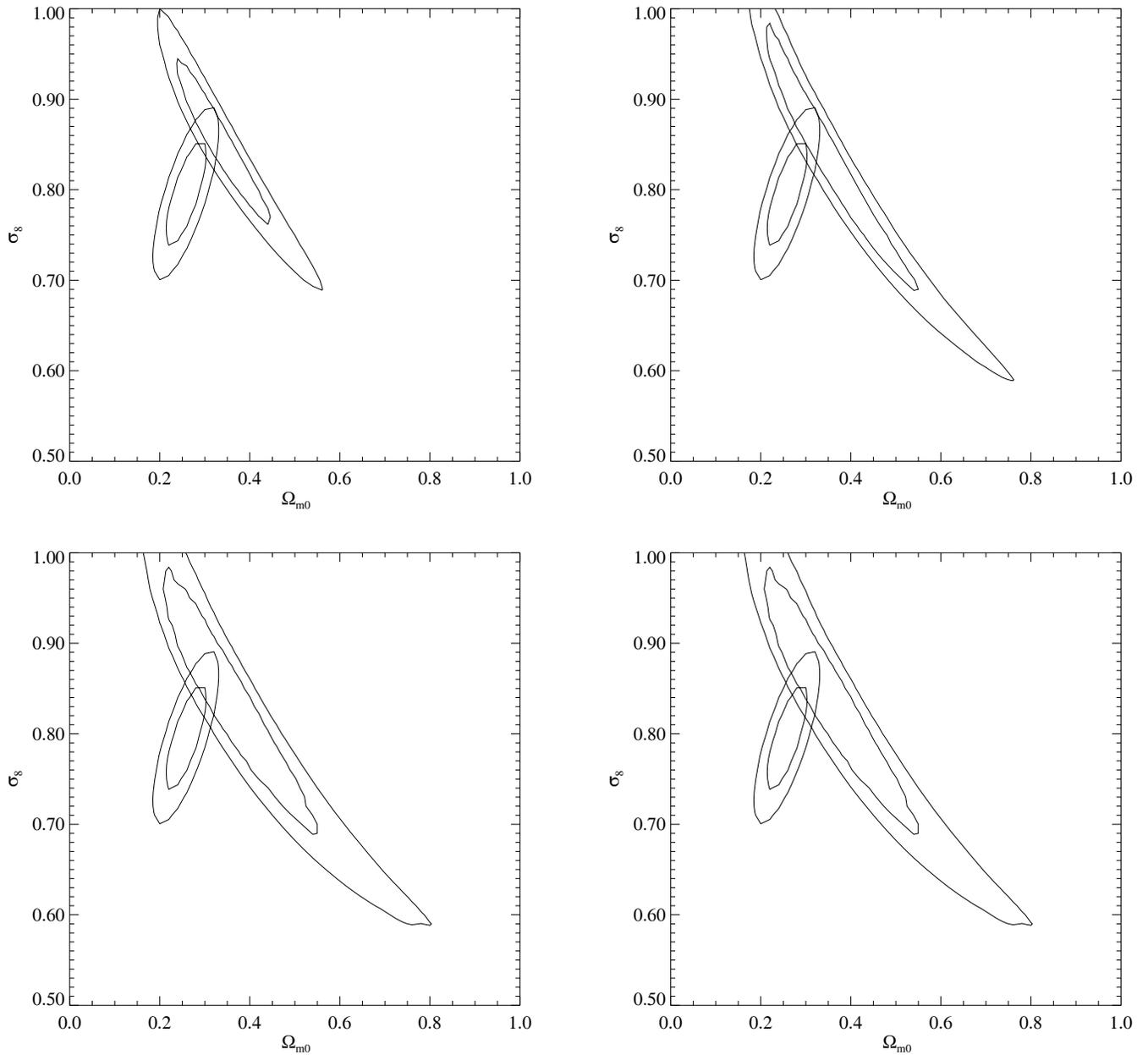

Fig. 10. Sixty-eight and ninety-five percent confidence contours for two parameters comparing the cluster temperature function constraints found here (long ellipses) to the WMAP5 constraints (short ellipses). The different panels show the result of marginalizing over increasingly more parameters. a. (*upper left*) statistical errors only. b. (*upper right*) Marginalizing over the L – T and M – T relations c. (*lower left*) Marginalizing over the L – T, M – T relations and the mass function. d. (*lower right*) Marginalizing over the L – T, M – T relations, the mass function, and h, $n_s$ and $\Omega_{b0} h^2$.

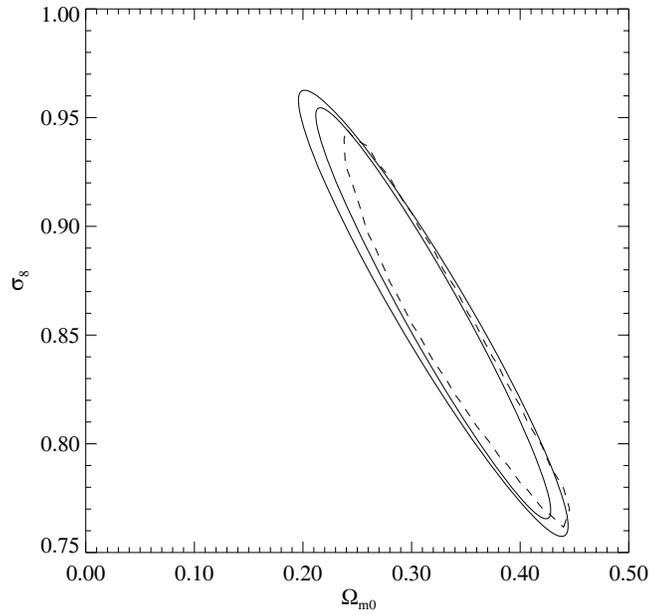

Fig. 12. Sixty-eight percent confidence error ellipses for two parameters comparing the actual Poisson statistical errors (dashed line) with the Fisher matrix Poisson and Poisson plus sample variance errors (solid lines).

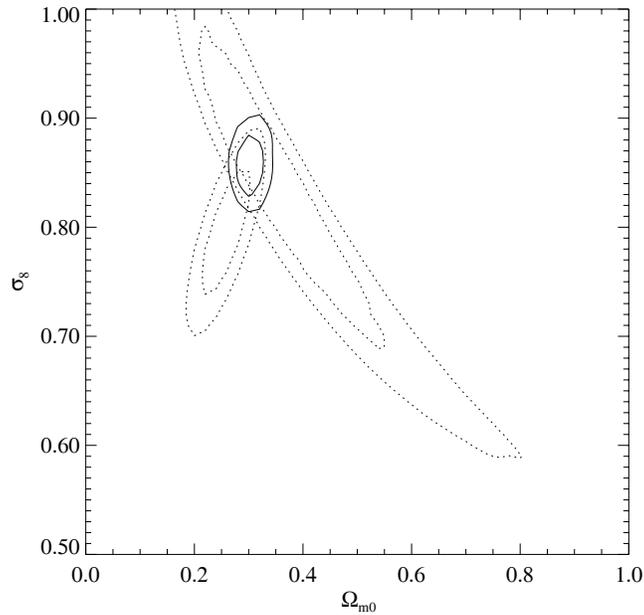

Fig. 13. Sixty-eight and ninety-five percent confidence contours for two parameters for the cluster temperature function constraints found here (dotted), the WMAP5 constraints (dotted) and the joint WMAP5 + cluster constraints (solid).

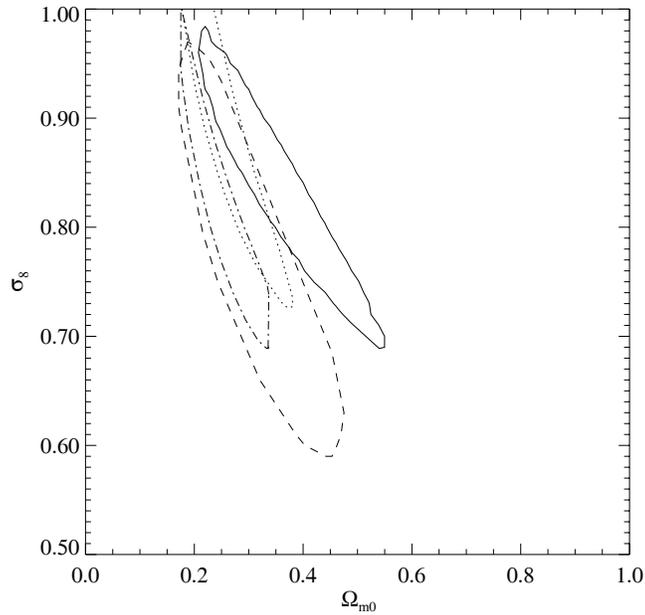

Fig. 14. Cosmological constraints derived from X-ray selected samples using different observables. The contours are at sixty-eight percent confidence for two parameters including systematic errors as estimated in the original work. Solid is using temperature (this work), dashed is using X-ray luminosity, dotted is using galaxy velocities and dash – dotted is using $Y_X$.

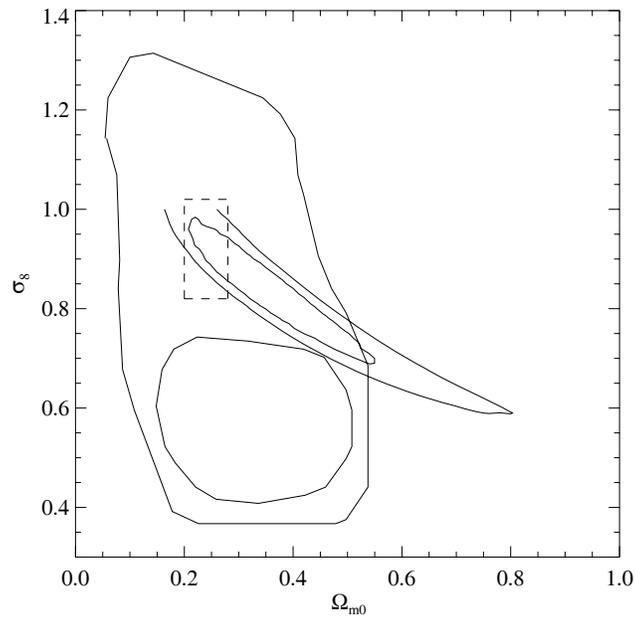

Fig. 15. Sixty-eight and ninety-five percent confidence contours for two parameters comparing the cluster temperature function constraints found here (long ellipses) to those from two optically selected samples; the RCS (circle and nearly vertical ellipse) and the SDSS maxBCG (dashed box, sixty-eight percent confidence for one parameter).

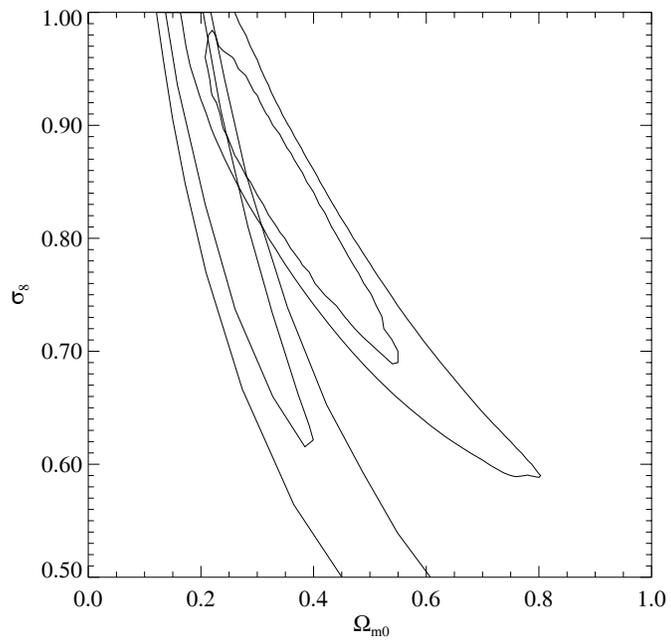

Fig. 16. Sixty-eight and ninety-five percent confidence contours for two parameters comparing the cluster temperature function constraints found here (upper contours) to the 100 Square Degree weak lensing shear constraints (lower contours).